# Design and optimization of a porous fence for the reduction of wind erosion

Adnan Tolga Kurumus[1], Nazhmiddin Nasyrlayev [2], Ender Demirel [3,*]

[1] Eskisehir Osmangazi University, Civil Engineering Department, 26480, Eskisehir, Turkey

[2] Computational Engineering for Sustainability Lab (CES-Lab), School of Engineering, University of Tasmania, Hobart, Australia

[3] Eskisehir Osmangazi University, Civil Engineering Department, 26480, Eskisehir, Turkey

* Correspondence: edemirel@ogu.edu.tr; Tel.: +90-222-2393750

ABSTRACT

A porous fence can be used as a shelter to reduce wind-induced erosion and the dispersion of dust particles from sand piles. Although many experimental and numerical studies have been carried out to analyze the effect of a porous fence on the erosion mechanism, optimizing porosity depending on in situ conditions remains a design challenge. This study focuses on the combined simulation and optimization of fence porosity considering the objective of reducing wind erosion over a triangular prism. Optimization of fence porosity is automated by integrating open-source computational fluid dynamics code and optimization tools to avoid trial and error. A comparison of numerical and experimental results demonstrates that the present numerical model can accurately predict turbulent flow through the porous fence for different porosities. Various fence shapes placed upstream of the prism were tested and optimized to reduce pressure and friction forces over the prism. The porosity of the new design is optimized according to the sand mitigation measure for different incoming wind speeds and sand grain diameters. The height of the fence is then optimized to prevent the onset of the erosion process under severe wind conditions. Unsteady three-dimensional simulation results are analyzed to reveal the underlying flow mechanism triggered by the porous fence. Eventually, the curved fence design proposed in the current study can be effectively used for the mitigation of wind erosion.

**Keywords:** Wind erosion; Wind fence; Sand movement; Simulation-optimization; Open-source.

Article Highlights

- A simulation-optimization framework is developed for the design optimization of porous fences.
- The novel fence design provides a significant reduction in erosion risk under severe wind conditions.
- The porosity and height of the novel fence are optimized according to the in situ conditions.
- Velocity and vorticity fields are investigated based on the unsteady three-dimensional simulations.

**Nomenclature**

| | | | |
|---|---|---|---|
| e | Distance between the fence and the triangular prism | $U_{ref}$ | Reference velocity |
| h | Prism height | $y_{ref}$ | Reference height |
| B | Prism half-width | K | von Kármán constant |
| $u_i$ | Velocity component in the $i^{th}$ direction | $v_{t\omega}$ | Turbulent viscosity |
| $x_i, x_j$ | Cartesian coordinates | $v_\omega$ | Kinematic viscosity near the wall |
| p | Pressure | y+ | Wall-normal distance parameter |
| $\rho$ | Air density | $C_p$ | Pressure coefficient |
| v | Air kinematic viscosity | $C_f$ | Friction coefficient |
| $S_i$ | Source term | $p_0$ | Reference pressure at the freestream |
| $\delta_{ij}$ | Kronecker delta function | $t_\omega$ | The mean wall-shear stress |
| k | Turbulence kinetic energy | $U_h$ | Velocity magnitude at the prism |
| $K_m$ | Turbulence kinematic viscosity | $F_1, F_2, F_3$ | Objective functions |
| $C_\mu$ | Eddy viscosity constant | l.b. ($\varphi$) | Lower bound of the design variable |
| $\varepsilon$ | Kinetic energy dissipation rate | u.b. ($\varphi$) | Upper bound of the design variable |
| $\sigma_k, \sigma_\varepsilon, C_{\varepsilon 1}, C_{\varepsilon 2}, \beta_0, \eta_0$ | Empirical constants of the RNG k-ε turbulence model | $N_c$ | Number of cells on the prism surface |
| R | Strain term | $\tau_x$ | Local shear stress |
| $\eta$ | Expansion parameter | $\tau_t$ | Threshold shear stress |
| $k_r$ | Pressure loss coefficient of the porous fence | d | Sand diameter |
| w | Thickness of the porous fence | SMM | Sand mitigation measure |
| $\varphi$ | Porous fence porosity | | |

1. **Introduction**

Extreme wind conditions resulting from the impact of climate change may cause significant morphological changes around engineering structures, such as transportation infrastructures, industrial facilities, farms, and buildings. The dispersion of heavy metals, originating from the erosion of tiny pieces of granular materials from aeolian landforms, can substantially contribute to air quality degradation in arid environments. Additionally, wind erosion of fertile topsoil in agricultural lands remains a major challenge for public authorities in the long term. Therefore, the risk assessment of wind erosion in engineering projects and the development of novel designs for mitigating adverse aerodynamic effects are gaining importance in the practical applications of wind engineering.

A vast number of research, including field studies [1-4], wind tunnel experiments [5-18], and numerical simulations [17-32], has been carried out to investigate wind erosion over piles of granular materials at industrial sites and to examine the turbulent flow structure altered by a wind shelter. The onset of the erosion process over aeolian landforms strongly depends on in situ conditions such as wind conditions, ground topography, sand diameter, and the shape and layout of the piles. The porous fence is distinguished among wind shelters due to its high performance in reducing wind effects. In a literature review of the aerodynamics and morphodynamics of sand fences, Li and Sherman [33] found that optimal fence designs were dependent on the purpose of the installation and the most significant factors controlling flow structure were the height and porosity of the wind fence. Additionally, experimental studies have demonstrated that adverse aerodynamic effects can be reduced using a porous fence, and the performance of the wind fence is mainly influenced by its porosity. A study by McClure, et al. [12] aimed to investigate the wind shelter effects of engineered porous fences designed using a typical multi-scale fractal structure, and to compare them to those of non-fractal fences. Two types of fractal fences, a two-dimensional (2D) and a one-dimensional (1D) fractal fence, were tested in a boundary-layer wind tunnel using a planar particle image velocimetry (PIV) method. Results showed that the 2D fractal fence had a clear favorable shelter effect, providing an extended sheltered area of 1.5-4 times the fence height and a shelter parameter of 0.4, about 70% longer than that of the non-fractal fence. Basnet and Constantinescu [21] conducted high-resolution three-dimensional (3D) large eddy simulations (LES) to investigate the flow physics past 2D solid and porous vertical plates mounted on a horizontal surface, with a fully developed turbulent incoming flow. The study examined the effects of the plate porosity, relative spacing between the solid elements of the porous plate, and roughness of the channel bed surface on the mean flow, turbulence structure, unsteady forces, and dynamics of the large-scale turbulent eddies. The simulation results revealed that for porosities less than 30%, the main recirculation eddy in the wake remained attached to the plate, while for larger porosities, it formed away from the porous plate. In a recent work, Dong, et al. [6] conducted wind tunnel tests to investigate the shelter effect of porous fences in a bulk terminal. The study considered three different arrangements of porous fences, eight incoming wind directions, and three different heights of fences. The results showed that the best wind reduction effect and economic efficiency were achieved with a combination of a 14 m high porous fence arranged perpendicular to the frequent wind direction. In one of the earliest studies in this area of research, Gandemer [8] demonstrated that the shelter effects in the lee of a fence could be estimated from the turbulent intensity and streamwise mean velocity. In a different study, it was reported that minor geometrical adjustments could lead to major enhancements in the protection of the arid area and sand trapping performance by decreasing friction velocity [20]. Guo, et al. [34] performed computational fluid dynamics (CFD) simulations to predict the pressure loss at a distance of eight perforation diameters downstream of the fence. The coefficient of the pressure loss increases when the inclination angle of the fence exceeds 30° and decreases when the surface roughness of the perforated plate increases. The performance of a porous fence with inclined openings was evaluated based on the unsteady Reynolds-averaged Navier-Stokes (URANS) simulations in order to assess the performance of the design in terms of mean turbulence quantities in the fence wake [35]. The streamwise velocity had less reduction when the bar angle decreased due to the increasing bleed flow, but the turbulence intensity decreased in the fence wake due to the increasing bleed flow. Deflector porous fence proposed by Chen et al., [22] could enlarge the protection area and considerably reduce the undesirable effects of the bleed flow on the windward side of the triangular prism. The literature indicates that porous fences have been extensively studied and implemented as wind shelters due to their abilities to reduce adverse aerodynamic effects. However, determining the optimal porosity of the wind fence remains a design challenge as it relies on atmospheric and topographic conditions to achieve the highest performance in reducing these effects. Previous studies demonstrate that the exploration of new fence designs is still of interest to researchers to attain higher shelter effects.

The growth of computational power has enabled researchers to conduct iterative numerical analyses and there is an increasing trend to optimize wind sheltering structures. The Reynolds-Stress model (RSM) was used [29] in conjunction with the gradient algorithm to optimize the porosity of the fence and to maximize the shelter effect in both the near wake region (0-4 $h_b$) and the far wake region (4 $h_b$ -10 $h_b$). It was found that a porosity of 10.2% provided the best shelter effect in the near wake region and a porosity of 22.1% in the far wake region. San et al. [30] used a gradient algorithm for the optimization of the porosity of a wind fence considering different objective functions such as integrated mean pressure, kinetic energy and peak velocity ratio. Restricted flow conditions and fence height were the principal limitations of the study. Horvat et al. [24] studied aerodynamic shape optimization to windblown sand barriers using the gradient-based optimization method based on the ratio of the wall shear stress and sand threshold shear stress. In their study on wind-blown sand protection of structures, Xin, et al. [16] investigated the optimum parameters of fence designs with perforated slats, using 3D CFD simulations to analyze the wind field and sand distribution around the fence. Subjective and objective evaluation methods were used to optimize the fence, and a porosity selection graph was proposed between the optimization parameters. Surrogate modelling optimization methods were used in some studies [27, 28] to improve the aerodynamic performance of porous fences. The results suggest that optimal fence height, porosity, and curved designs can lead to substantial improvements in the sheltering effects of porous fences.

Although many new studies on the optimization of porous wind fences are emerging, the current solution approach is primarily based on two-dimensional numerical analyses and surrogate-based optimization methods. However, the aerodynamic effects governing wind shelters are complex and include the unsteady nature of atmospheric turbulence and the three-dimensional characteristics of vortex structures. These high Reynolds number phenomena may not be accurately captured using two-dimensional Reynolds-averaged Navier-Stokes (RANS) methods [36]. Additionally, while surrogate-based optimization methods have shown promise in previous studies, they may require significant computational resources and introduce errors and uncertainties in the optimization results. Therefore, in this study, we will primarily rely on two-dimensional numerical analyses coupled with two gradient-based and one derivative-free optimization methods to construct an automated simulation and optimization framework. For optimized cases, we will utilize an unsteady three-dimensional numerical analysis to better understand the complex flow structures and high Reynolds number effects. This combination of two and three-dimensional analyses was chosen to shorten the amount of time spent on computation and lower the computational cost.

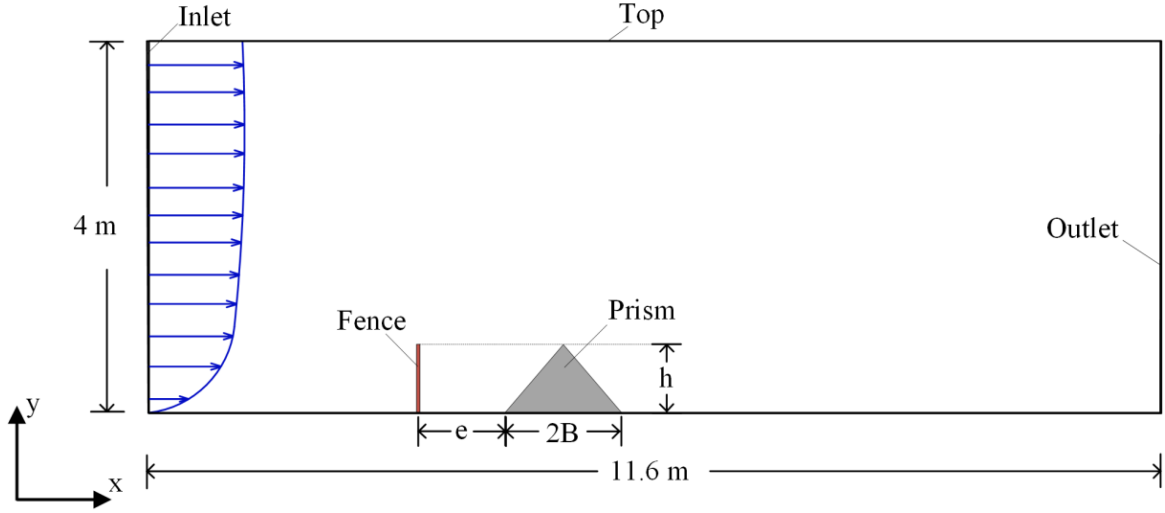

**Fig. 1**. Computational domain and boundary conditions for the ABL flow through a prism with a fence.

A computational domain with a porous fence located upstream of a triangular prism (e = 0.06 m) is shown in Fig. 1. The purpose of the fence is to mitigate adverse aerodynamic effects acting on the prism. The height of the prism is h = 0.04 m, and the base width is 2B = 0.095 m. The computational domain length is 290h, and the height is 100h. An ABL develops through the prism, and the flow emerges from the outlet of the domain with no reflection. The reliability of the present computational model was confirmed by comparing numerical results with experimental measurements from the literature. Building on the validated model, an automatic optimization framework was developed using an open-source CFD code and an optimization tool. The framework was used to optimize the porosity and height of the wind fence located upstream of the triangular prism. Various fence shapes were tested and optimized to reduce pressure and friction forces over the surface. The porosity of the new design was optimized according to the sand mitigation measure for different incoming wind speeds and sand grain diameters, while the fence height was optimized to prevent erosion under severe wind conditions. Additionally, open-source cases were publicly distributed for transparency of the results. This work aims to contribute to the field of porous fence optimization through the use of automated optimization methods, avoiding the need for intuitive estimations of porosity for different fence geometries and atmospheric conditions.

## 2. Computational model and validation

*2.1. Numerical model*

The continuity and momentum equations that describe incompressible turbulent flow through the wind fence can be expressed using Reynolds averaging of the instantaneous velocity fluctuations for steady-state flow as:

$$\frac{\partial u_i}{\partial x_i} = 0 \qquad (1)$$

$$u_j \frac{\partial u_i}{\partial x_j} = -\frac{1}{\rho}\frac{\partial p}{\partial x_i} + \frac{\partial}{\partial x_j}\left(\nu \frac{\partial u_i}{\partial x_j} - \overline{u_i' u_j'}\right) + S_i \qquad (2)$$

where, $u_i$ is the $i$ th direction velocity component, $p$ is the pressure, $\rho$ is the air density, $v$ is the kinematic viscosity of the air and $S_i$ is the source term to account for the resistance by the porous fence. Reynolds stresses can be calculated from the following Boussinesq approximation:

$$-\overline{u_i' u_j'} = K_m \left( \frac{\partial u_i}{\partial x_j} + \frac{\partial u_j}{\partial x_i} \right) - \frac{2}{3} \delta_{ij} k \tag{3}$$

where, $\delta_{ij}$ denotes the Kronecker delta function, $k$ is the turbulence kinetic energy and $K_m$ is the turbulence kinematic viscosity, which can be calculated from the following equation for the Re-Normalization Group (RNG) k-ε turbulence closure model:

$$K_m = v \left( 1 + \left( \frac{C_\mu}{v} \right)^{1/2} \frac{k}{\varepsilon^{1/2}} \right)^2 \tag{4}$$

where, $C_\mu = 0.09$ is an empirical constant and $\varepsilon$ is the kinetic energy dissipation rate. The turbulence kinetic energy and dissipation rate can be calculated using the transport equations in the RNG k-ε turbulence model as follows:

$$u_j \frac{\partial k}{\partial x_j} = -\overline{u_i' u_j'} \frac{\partial u_i}{\partial x_j} + \frac{\partial}{\partial x_i} \left( \frac{K_m \partial k}{\sigma_k \partial x_j} \right) - \varepsilon \tag{5}$$

$$u_j \frac{\partial \varepsilon}{\partial x_j} = -C_{\varepsilon 1} \frac{\varepsilon}{k} \overline{u_i' u_j'} \frac{\partial u_i}{\partial x_j} + \frac{\partial}{\partial x_j} \left( \frac{K_m \partial \varepsilon}{\sigma_\varepsilon \partial x_j} \right) - C_{\varepsilon 2} \frac{\varepsilon^2}{k} - R \tag{6}$$

Empirical constants of the RNG k-ε turbulence model are $\sigma_k = 0.7179$, $\sigma_\varepsilon = 0.7179$, $C_{\varepsilon 1} = 1.42$, $C_{\varepsilon 2} = 1.68$, $\beta_0 = 0.012$, $\eta_0 = 4.377$ [37, 38]. The extra strain term $R$ is expressed as follows:

$$R = \frac{C_\mu \eta^3 \left(1 - \frac{\eta}{\eta_0}\right) \varepsilon^2}{(1 + \beta_0 \eta^3) k} \tag{7}$$

$$\eta = \frac{k}{\varepsilon} \left[ \left( \frac{\partial u_i}{\partial x_j} + \frac{\partial u_j}{\partial x_i} \right) \frac{\partial u_i}{\partial x_j} \right]^{1/2} \tag{8}$$

The source term in Eq. (2) can be modeled as a porous jump to account for the pressure drop along the porous fence [39]:

$$S_i = C_2 \frac{1}{2} \rho |u| u_i \tag{9}$$

where, $C_2 = k_r/w$, $k_r$ is the pressure loss coefficient of the fence and $w$ is the thickness of the fence as 1 mm [14]. Reynolds [40] proposed the pressure loss coefficient as $k_r = 1.04(1 - \phi^2)/\phi^2$, where $\phi$ denotes the fence porosity.

## 2.2. Mesh and boundary conditions

In the present study, a block-structured mesh is generated using the blockMesh utility in OpenFOAM for the two-dimensional domain. A two-dimensional computational model is adopted because preliminary simulations showed that two and three-dimensional results were identical. As shown in Fig. 2, the mesh is clustered near the bottom, prism crest, and wind fence to capture severe variations in the flow structure. Additionally, the mesh is refined around the fence to accurately account for the resistance through the fence.

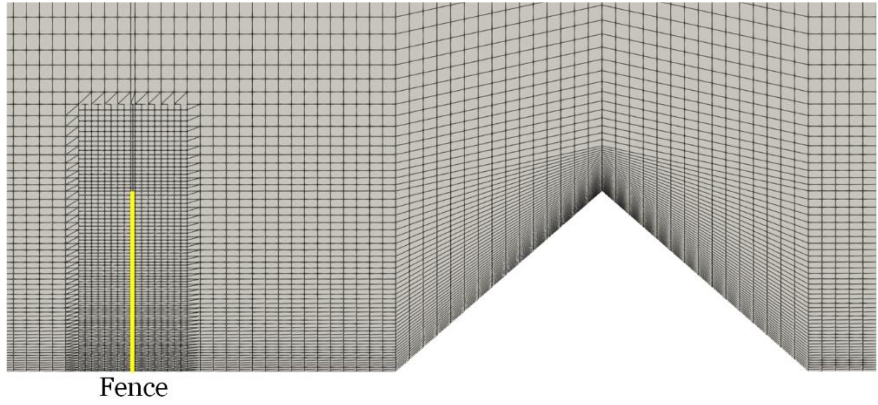
Fence

**Fig. 2**. Block-structured mesh for the flow through a wind fence and triangular prism.

The following boundary conditions are used to impose an ABL at the inlet of the domain [41]:

$$U = \frac{u_*}{K} \ln\left(\frac{y+y_0}{y_0}\right) \quad (10)$$

$$k = \frac{u_*^2}{\sqrt{C_\mu}} \quad (11)$$

$$\varepsilon = \frac{u_*^3}{K(y+y_0)} \quad (12)$$

$$u_* = \frac{K U_{ref}}{\ln\left(\frac{y_{ref}+y_0}{y_0}\right)} \quad (13)$$

where, $U_{ref}$ and $y_{ref}$ are reference velocity and reference height, respectively, and $K = 0.41$ is the von Kármán constant. The $U_{ref}$ is equal to the freestream velocity, which is set to 14.5 m/s and the reference height is $y_{ref} = 0.18$ m according to the experimental studies conducted by Park and Lee [14]. The turbulence viscosity is calculated from the following equation in the ABL flow:

$$\nu_{t\omega} = \nu_\omega \left(\frac{y^+ K}{\ln\left(\max\left(\frac{y+y_0}{y_0}, 1+10^{-4}\right)\right)} - 1\right) \quad (14)$$

where, $\nu_{t\omega}$ is turbulent viscosity, $\nu_\omega$ is the kinematic viscosity near the wall and $y^+ = u_* y/\nu_\omega$. Empty boundary conditions were used at the lateral sides of the domain to reduce the three-dimensional numerical model to a two-dimensional model. Thus, the flow is solved in the *x-y* domain shown in Fig. 1. Boundary conditions of the problem are given in the following table in detail.

**Table 1**

Boundary conditions for the turbulent flow through a porous fence.

| Variable | Location | Type | Boundary condition | Value |
|---|---|---|---|---|
| U, k, epsilon | inlet | patch | atmBoundaryLayerInlet | Eqn. (10), (11) and (12) |
| U, k, epsilon | outlet | patch | zeroGradient | |
| U | top | patch | fixedShearStress | |
| k, epsilon, p, nut | top | patch | zeroGradient | |
| p | inlet | patch | zeroGradient | |
| p | outlet | patch | uniformFixedValue | uniform 0 |
| nut | inlet, outlet, top | patch | calculated | uniform 0 |
| U | bottom, prism | wall | noSlip | |
| k | bottom, prism | wall | kqRWallFunction | |
| epsilon | bottom, prism | wall | epsilonWallFunction | uniform 0 |
| p | bottom, prism | wall | zeroGradient | uniform 0 |
| nut | bottom, prism | wall | atmNutkWallFunction | Eqn. (14) |
| U, k, epsilon, p, nut | lateral | empty | empty | |

The semi-implicit method for pressure-linked equations (SIMPLE) solver was used to solve the pressure-velocity coupling problem. Solutions were considered converged to the steady-state when the residuals were below $1x10^{-5}$ for all flow variables. Gradient and Laplacian terms were discretized using the second-order Gauss linear method. Variables were calculated at cell faces using a second-order linear method to calculate convection and diffusion fluxes. Convective terms in the momentum and transport equations were discretized using second-order accurate Gauss linear and Gauss limiter linear methods, respectively, to overcome stability issues related to turbulence fluctuations. Equations were discretized using second-order accurate numerical schemes, and corresponding coefficients were stored into matrices for the implicit solution of the governing equations. The GAMG (Geometric-Algebraic Multi-Grid) method was used in the solution of the linear system for pressure, and the smooth method was used in the solution of the linear system for velocity, $k$ and $\varepsilon$. The implicit solution of the governing equations increases the stability of the numerical model compared to explicit methods.

*2.3. Mesh independence study*

A mesh sensitivity study was carried out to find the optimum number of cells and reduce the total simulation duration during optimization studies. Different mesh resolutions were created, and the corresponding minimum, average, and maximum $y^+$ values are given in Table 2. The wall functions in Table 1 are unified wall functions that account for the wall effects when the dimensionless wall distance is in the viscous sublayer or transition zone.

**Table 2**

Mesh features and $y^+$ values near the boundaries.

| Mesh | Cell number | $y^+_{min}$ | | $y^+_{avg}$ | | $y^+_{max}$ | |
| --- | --- | --- | --- | --- | --- | --- | --- |
| | | Bottom Wall | Prism | Bottom Wall | Prism | Bottom Wall | Prism |
| 1 | 97050 | 13.10 | 4.62 | 67.30 | 22.32 | 71.46 | 42.59 |
| 2 | 135870 | 5.91 | 0.71 | 21.73 | 6.96 | 23.65 | 14.04 |
| 3 | 174690 | 3.10 | 0.304 | 14.40 | 3.48 | 15.76 | 8.32 |
| 4 | 210128 | 2.27 | 0.20 | 10.65 | 2.38 | 12.30 | 5.74 |
| 5 | 232920 | 2.06 | 0.14 | 10.62 | 2.30 | 11.76 | 5.69 |

The performance of the numerical model is evaluated based on the pressure and friction coefficients presented below:

$$C_p = \frac{p - p_0}{\frac{1}{2}\rho U_h^2} \tag{15}$$

$$C_f = \frac{\tau_w}{\frac{1}{2}\rho U_h^2} \tag{16}$$

Here, $p_0$ represents the reference pressure at the freestream, $p$ is the mean surface pressure, $\tau_w$ is the mean wall-shear stress, the denominator represents the mean dynamic pressure at the crest of the triangular prism and the $U_h$ is the velocity magnitude at the prism.

Variations in the pressure and friction coefficients are plotted in Fig. 3 for the mesh resolutions given in Table 2 to determine the number of cells at which normal and shear stresses remain unchanged. The differences observed in the pressure and friction coefficients become negligible when the mesh resolution is higher than that in Mesh 4. Mesh 5 is used for the remaining simulations in the present study to ensure accuracy.

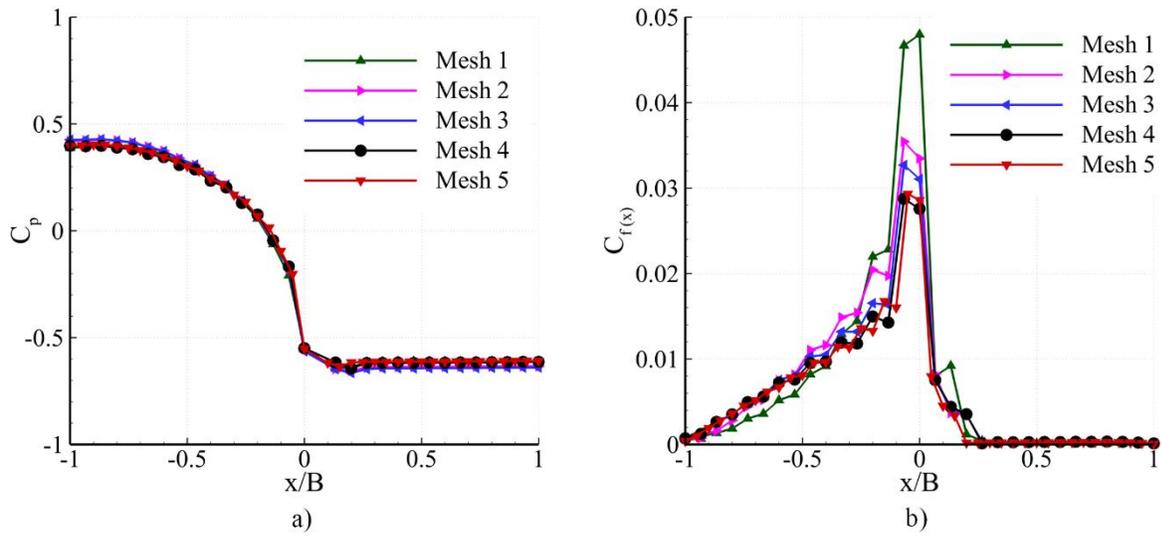

**Fig. 3.** Comparison of a) pressure and b) friction coefficients along the prism for different mesh resolutions.

Fig. 4 shows the variations of $y^+$ and friction coefficient along the bottom boundary for Mesh 5. The abrupt increase observed on the prism in Fig. 4b is due to the accelerating flow and increasing wall-shear stress in this region. The shear stress distribution in Fig. 4b indicates the equilibrium ABL at the inlet and far end of the computational domain. Therefore, the present inlet and outlet boundary conditions maintain an ABL for the turbulent flow through the fence.

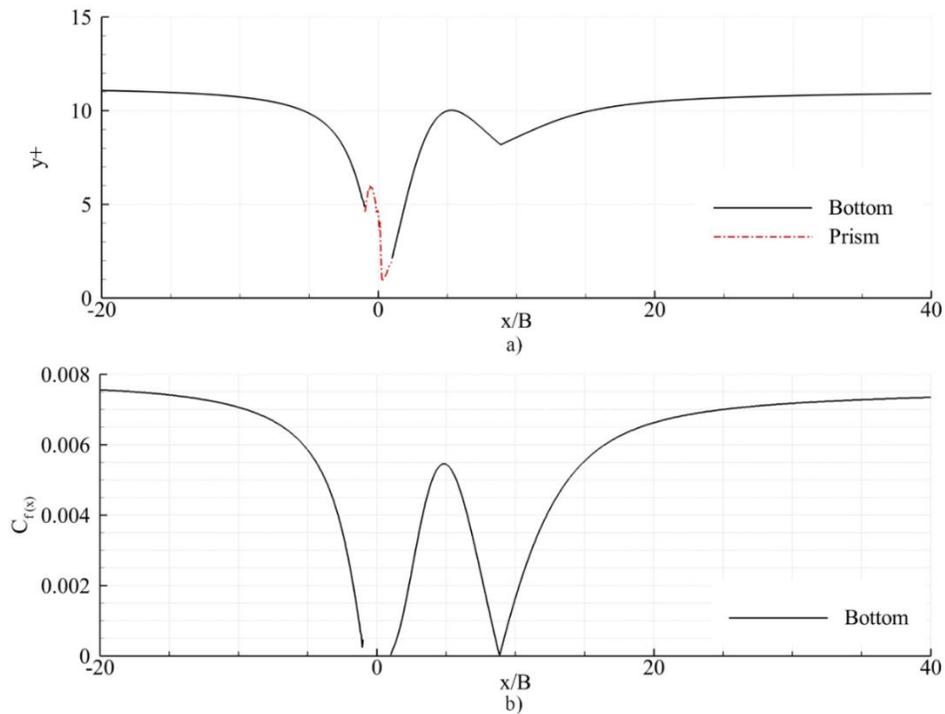

**Fig. 4.** Variations of a) $y^+$ and b) friction coefficient along the bottom.

*2.4. Validation of the numerical model*

Numerical simulations were performed under the same flow conditions as in the experimental study conducted by Park and Lee [14]. The numerical simulation results were compared with the experimental measurements for the cases of no fence and a porous fence, as shown in Fig. 5. As seen in the figure, the maximum $C_p$ was observed upstream of the prism, and the pressure coefficient dropped abruptly near the crest of the prism. The porous fence successfully reduced pressure fluctuations, and $C_p$ tended to decrease on the windward surface of the prism as the porosity of the fence decreased, whereas $C_p$ values were observed to be almost unchanged on the leeward surface of the prism. This observation indicates that the flow separation is almost the same on the leeward surface of the prism, and the upstream face of the prism requires the development of an efficient shelter design. However, an optimization procedure needs to be performed through a series of simulations to find the optimized design in terms of the reductions in normal and shear stresses acting on the prism. We propose a simulation-optimization framework, in which design parameters are updated automatically during numerical simulations to determine the minimum value of an objective function depending on design purposes.

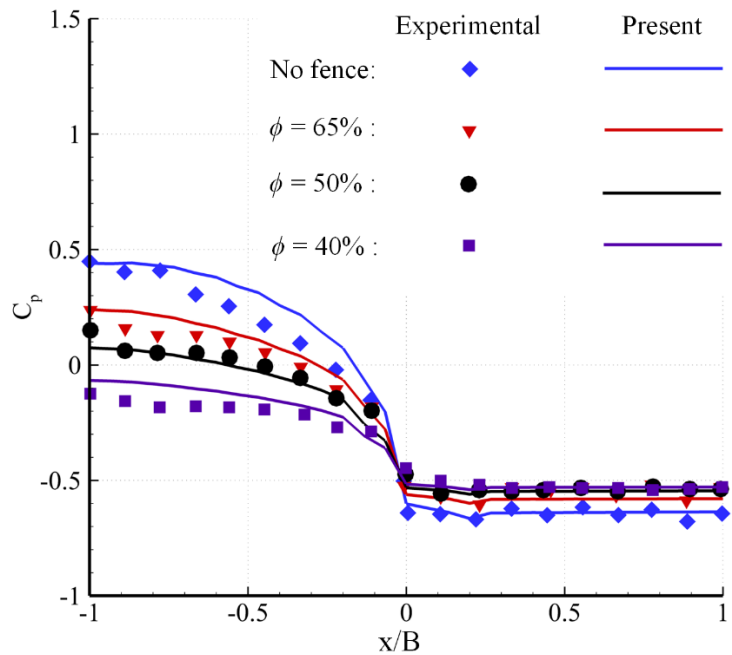

**Fig. 5**. Comparison of the simulated pressure coefficients with the experimental data.

*2.5. Optimization of the porosity*

Coupling an optimization tool with a CFD solver may provide design automation for the optimization of an engineering design once design parameters and objective functions are defined. Based on this idea, the present study aims to optimize the porosity of a porous fence to reduce wind effects on a triangular prism by coupling an optimization tool with a CFD solver. Open-source tools OpenFOAM and Dakota are used for the CFD simulation of turbulent flow through the fence and optimization of the fence porosity, respectively. Python and bash scripts are used to integrate the CFD code and optimization tool and to provide a simulation-optimization framework for optimizing the fence porosity. The proposed framework and data can be used to optimize the design of an engineering system by defining design parameters to be optimized based on objective functions.

Optimization is performed by iteratively updating the design parameters using a search method that aims to find a more accurate value for the objective function than the current one. In this study, the Fletcher-Reeves conjugate gradient (FRCG), the full Newton, and the division of rectangles (DIRECT) methods were used as optimization algorithms in Dakota. The Fletcher-Reeves conjugate gradient and full Newton methods search for the optimal design parameters by calculating the gradient of the response function to determine the direction of iteration, based on a gradient-based optimization algorithm. On the other hand, the DIRECT method is a derivative-free global method that subdivides the design space into smaller rectangles over a finite number of iterations [42].

Dakota optimization software was coupled with OpenFOAM to optimize the porosity and height of the fence depending on the ABL conditions. The present optimization problem can be formulated in the following generic form:

$$Minimize: F_1(\phi), F_2(\phi) \ldots F_N(\phi)$$

$$s.t. \; l.b.(\phi) < \phi < u.b.(\phi) \tag{17}$$

where, $F_1(\phi)$ and $F_2(\phi)$ are the objective functions subject to lower $l.b.(\phi)$ and upper $u.b.(\phi)$ bounds of the design variable, $\phi$ is the design variable and $N$ is the number of objective functions. Large pressure fluctuations are not anticipated in the present problem to reduce particle drift of coal stockpiles [11]. The following objective functions are constructed by integrating absolute values of the pressure and friction coefficients on the prism surface:

$$Minimize: F_1(\phi) = \sum_{i=1}^{N_c} |C_{pi}(\phi)| \tag{18}$$

$$Minimize: F_2(\phi) = \sum_{i=1}^{N_c} |C_{fi}(\phi)| \tag{19}$$

where, $C_{pi}(\phi)$ and $C_{fi}(\phi)$ are the pressure and friction coefficients at the $i^{th}$ cell on the prism surface as discrete values, and $N_c$ is the number of cells on the surface.

The simulation and optimization algorithm developed in the present study are performed as shown in Fig. 6. The Dakota software first initializes the design parameters and passes them to the OpenFOAM solver. The CFD solver simulates the flow using the design parameters, and the results are post-processed by scripts to calculate the pressure ($C_p$) and friction ($C_f$) coefficients using Eqs. (15) and (16), respectively. These results are evaluated using Eqs. (18) and (19) and returned to the Dakota software to set the porosity for the next step. The optimization cycle is completed, and the optimal design parameters are achieved when Eqs (18) and (19) are satisfied.

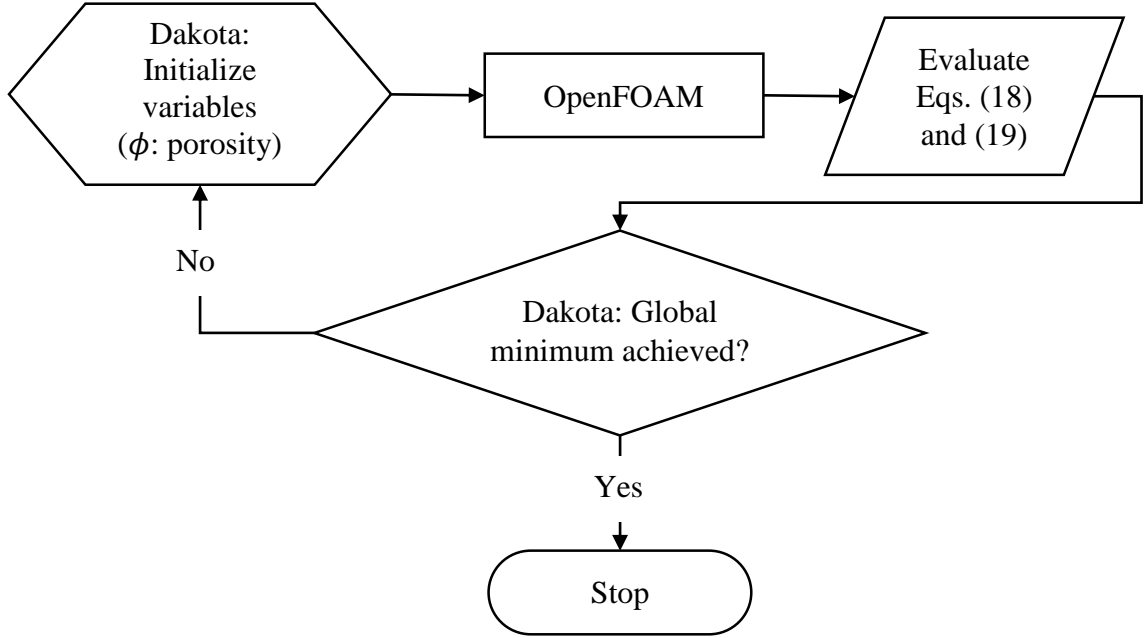

**Fig. 6**. Flow chart for the simulation and optimization framework.

## 3. Results and discussions

### 3.1. Performance of the optimization methods

In this study, objective optimization is accomplished using three optimization algorithms in Dakota. Fig. 7a shows the variations of the integrated pressure coefficient with porosity for the selected optimization algorithms. The current simulation-optimization algorithm aims to minimize the objective function by varying the porosity. The optimal porosity was found to be 46.36%, resulting in a minimum absolute total pressure coefficient of $\sum_{i=1}^{N_c}|C_{pi}| = 9.775$ over the triangular prism consisting of 30 cells, or a mean pressure coefficient of $\overline{|C_p|}= 0.327$. The optimization process is initiated by the Dakota optimization software, and all processes are performed successively by the proposed simulation-optimization framework without any interruption by user until the optimization is achieved. The variation of the integrated friction coefficient with porosity is shown in Fig. 7b. The friction coefficient increases as the porosity increases due to the decreasing velocity blockage and reaches its minimum value for a small porosity of 13.35%.

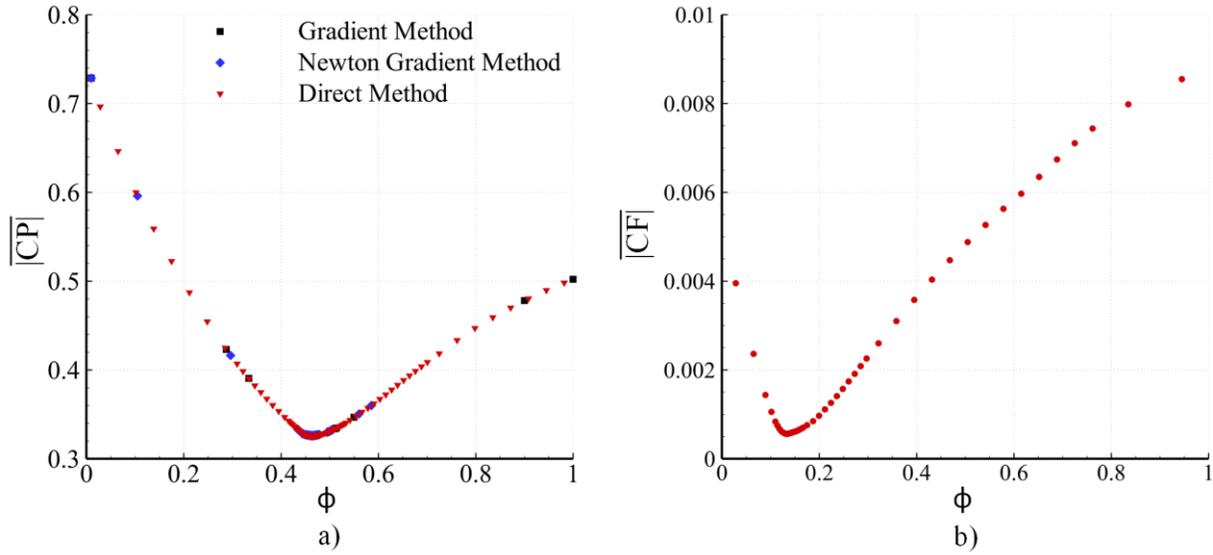

**Fig. 7**. Performance assessment of the optimization algorithms. (a) Optimization of the pressure coefficient with various methods and (b) optimization of the friction coefficient with Direct Method.

The plots of pressure and friction optimization results in Fig. 7 demonstrate that the optimization process is well-behaved. This means that all the optimization methods quickly converge to the optimum. The reason for this behavior can be attributed to the one-dimensional nature of the optimization problem, where the only design parameter is porosity. The convergence speeds of the optimization algorithms for $C_p$ optimization are compared in Table 3, which includes the number of iterations and CPU time required for each method. Efficiency can be measured in terms of both the number of iterations and CPU time required to converge to a solution. Among the gradient methods, the FRCG method took fewer iterations and thus less CPU time, making it the most efficient method, requiring only 27 iterations and 5,357.5 seconds of CPU time. The Full Newton method requires more iterations (54) and double the CPU time (10,727.7 seconds) than the FRCG method. The DIRECT method requires even more iterations (117) and significantly more CPU time (23,354.4 seconds) than the other two methods. This is expected, as non-gradient-based optimization algorithms tend to consider a larger number of design spaces and require more function evaluations to find the optimal solution. To reduce the total CPU time, simulations were conducted using 56 processors with parallel computing strategies on the High-Performance and Grid Computing Center (TRUBA).

**Table 3.** Comparison of optimization algorithms for the $C_p$ optimization: Iteration numbers and CPU times.

| Method | Iteration Number | CPU Time (s) |
| --- | --- | --- |
| FRCG | 27 | 5,357.5 |
| Full Newton | 54 | 10,727.7 |
| DIRECT | 117 | 23,354.4 |

It is worth noting that the error between different optimization algorithms is not directly comparable since an exact solution is not available. The error can be defined as the difference between the optimized solution and the true optimal solution. Nevertheless, a comparison of the optimum porosities for the porous fence located upstream of the triangular fence was made with the results of prior published literature, as shown in Table 4. Note that all three optimization algorithms in the present study converged to the same optimum solution, indicating a negligible discrepancy ($\approx 0\%$) between them for the same dimensions of domain as in the reference case used in the validation part. The optimization algorithms in the present study yielded slightly better results with an optimum porosity of 46% and an absolute mean pressure coefficient of 0.324, compared to previous studies that reported optimum porosities of 40% and 51.9% and absolute mean pressure coefficients of 0.368 and 0.34, respectively. Since the study by Park and Lee [14] involved experimental analysis rather than numerical optimization, it is not reasonable to calculate the discrepancy between their study and the present one. However, the study by San, et al. [30] involved numerical optimization, and hence the difference can be calculated between their study and the present one. This difference is calculated as $\left(\overline{|C_p|}_1 - \overline{|C_p|}_2\right)/\overline{|C_p|}_2 \times 100\%$, where $\overline{|C_p|}_1$ and $\overline{|C_p|}_2$ correspond to the absolute mean pressure coefficients of the present and compared studies over the prism surface, respectively. The calculated difference of 3.6% suggests that these results are consistent with each other with a minimal difference and porosity values of both studies are quite close.

**Table 4.** Comparison of the optimum porosities with respect to the pressure coefficient.

| Reference | Optimum porosity (%) | $\overline{|C_p|}$ |
|---|---|---|
| Lee and Park [11], Park and Lee [14] | 40% | 0.368 |
| San, et al. [30] | 51.9% | 0.340 |
| Present study | 46.36% | 0.327 |

Fig. 8 depicts variations of the integrated pressure coefficients over the prism surface for no fence and optimized fence. The optimized porous fence could substantially reduce the pressure coefficient over the prism and can be used upstream of the prism as an effective wind shelter.

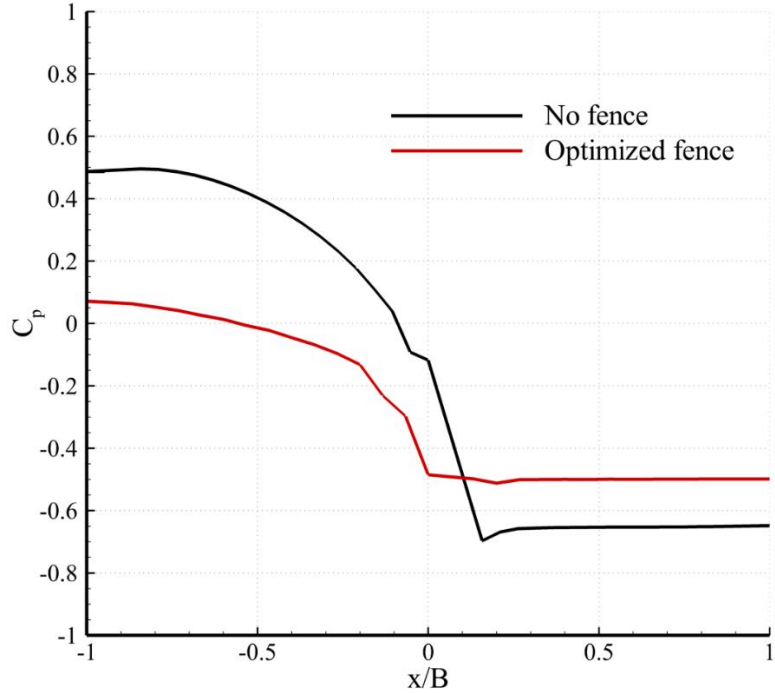

**Fig. 8**. Distributions of the pressure coefficients over the prism for the optimized design ($\phi = 0.46$) and no fence.

*3.2. Development and optimization of a novel fence design*

Various fence designs (Fig. 9) are considered to investigate the effect of fence geometry on the shelter performance. The heights of the fences are set to the same height as in the conventional design, and the porosity is selected as 50%, which is close to the optimal porosity of the conventional design for pressure coefficient to enable systematic comparison of results. Each fence is located at the same position as in the conventional design, and deviations from the vertical lines are 8 mm for the shapes in Fig. 9, except for design 11. The present numerical model can test different shelter designs, and the performance of each design can be assessed according to the objective functions to achieve the highest shelter performance. Thus, numerical simulations are performed for each design, and the performance enhancement achieved in comparison to the conventional design is given in Table 5 in terms of pressure and friction effects.

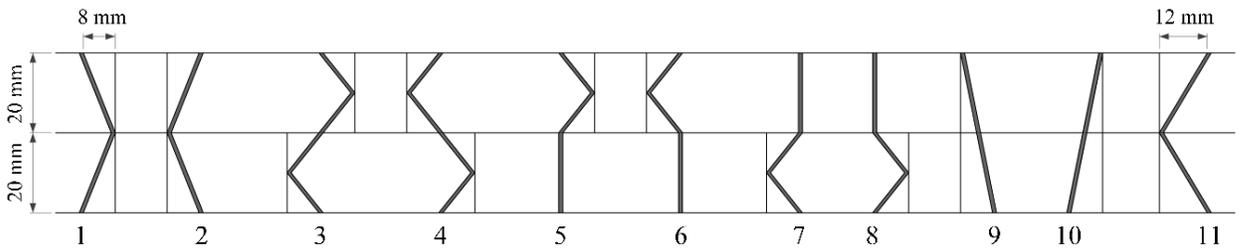

**Fig. 9**. Various fence geometries tested for performance improvement.

Streamwise arrow-shaped designs 1, 3, 5, and 8 led to an increase in the friction effects on the prism surface. However, design 2 could enhance both pressure and friction reductions due to its arrow shape pointing against the stream vectors. A comparison of the results of designs 4 and 6 indicates that using shapes that inclined against the wind direction on the upper half of the fence resulted in reducing pressure forces on the prism. However, using those shapes on the lower parts of designs 3 and 7 resulted in degrading performance on both pressure and friction coefficients. The results of design 9 and design 10, as well as design 2, provide insights into the development of design 11 since the inclination angle has a trade-off improvement in terms of pressure and friction effects.

**Table 5**

Improvements in pressure and friction coefficients in comparison to the conventional design.

| Design | $\overline{[C_f]}$ | Enhancement (%) | $\overline{[C_p]}$ | Enhancement (%) |
| --- | --- | --- | --- | --- |
| Conventional %50 | 0.0048 | - | 0.329 | - |
| 1 | 0.00516 | -7.67 | 0.333 | -1.1 |
| 2 | 0.00452 | 5.83 | 0.324 | 1.57 |
| 3 | 0.0052 | -8.43 | 0.335 | -1.67 |
| 4 | 0.00467 | 2.58 | 0.324 | 1.42 |
| 5 | 0.00513 | -7.06 | 0.334 | -1.31 |
| 6 | 0.0046 | 4.07 | 0.323 | 1.95 |
| 7 | 0.00488 | -1.81 | 0.328 | 0.20 |
| 8 | 0.00481 | -0.24 | 0.329 | 0.09 |
| 9 | 0.00469 | 2.14 | 0.327 | 0.65 |
| 10 | 0.00491 | -2.35 | 0.331 | -0.70 |
| 11 | 0.00442 | 7.89 | 0.322 | 2.25 |

It is expected that the novel design will decrease both pressure and friction effects. Numerical results in Table 5 show that design 11 outperforms the conventional fence design in terms of pressure and friction forces. Therefore, optimization studies are conducted for design 11 to achieve the highest shelter performance. Reductions in pressure and friction coefficients are compared in Fig. 10 for the conventional and new designs. The optimizations of pressure and friction coefficients follow the same trend after 46% porosity. The total reduction in design 11 is calculated as the summation of pressure and friction reductions, using equal weight, and is compared in Fig. 10b. A designer can conduct trade-off analysis using different weights of reductions in Fig. 10a for design purposes.

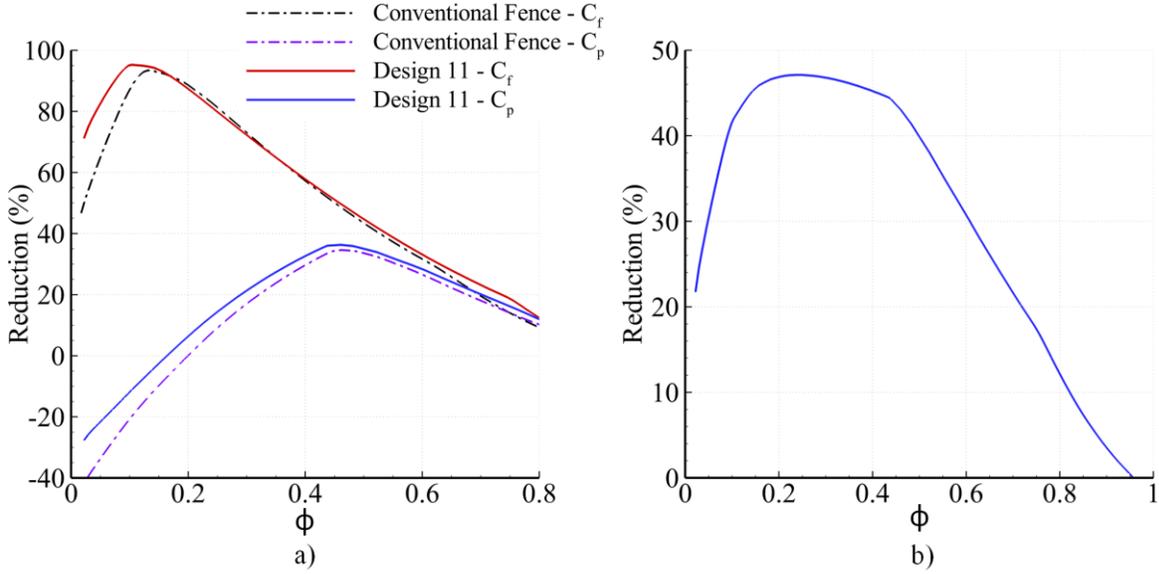

**Fig. 10.** Variation of the weighted reduction with the porosity: a) Optimizations of $C_f$ and $C_p$ for conventional design and design 11 relative to the no fence case and b) variation of total reduction for different porosities of design 11 using equal weights.

Distributions of velocity magnitude and velocity vectors are visualized in Fig. 11 for each design to understand the underlying mechanics of reducing both pressure and friction forces on the prism. The maximum velocity is observed around the fence, and the separated flow accelerates above the prism, increasing the shelter performance. While accelerating flow at the windward face of the prism may increase friction forces on the prism due to the low fence height, using a higher fence might reduce friction forces but also increases construction costs. A recirculating zone forms on the leeward side of the prism, reducing both drag and lift coefficients. The presence of the fence causes high-flow parcels to move away from the leeward face of the prism and increases the shelter performance. If the leeward face of the prism requires additional protection from excessive vacuum pressure, an additional fence can be located in the leeward zone to diminish the recirculation zone. Additionally, the outward gap at the bottom half of the design induces low pressure and velocity magnitudes by splitting and absorbing wind energy. This feature of the design is mainly beneficial to the windward side of the prism, which is the most crucial surface in terms of cavitation risks.

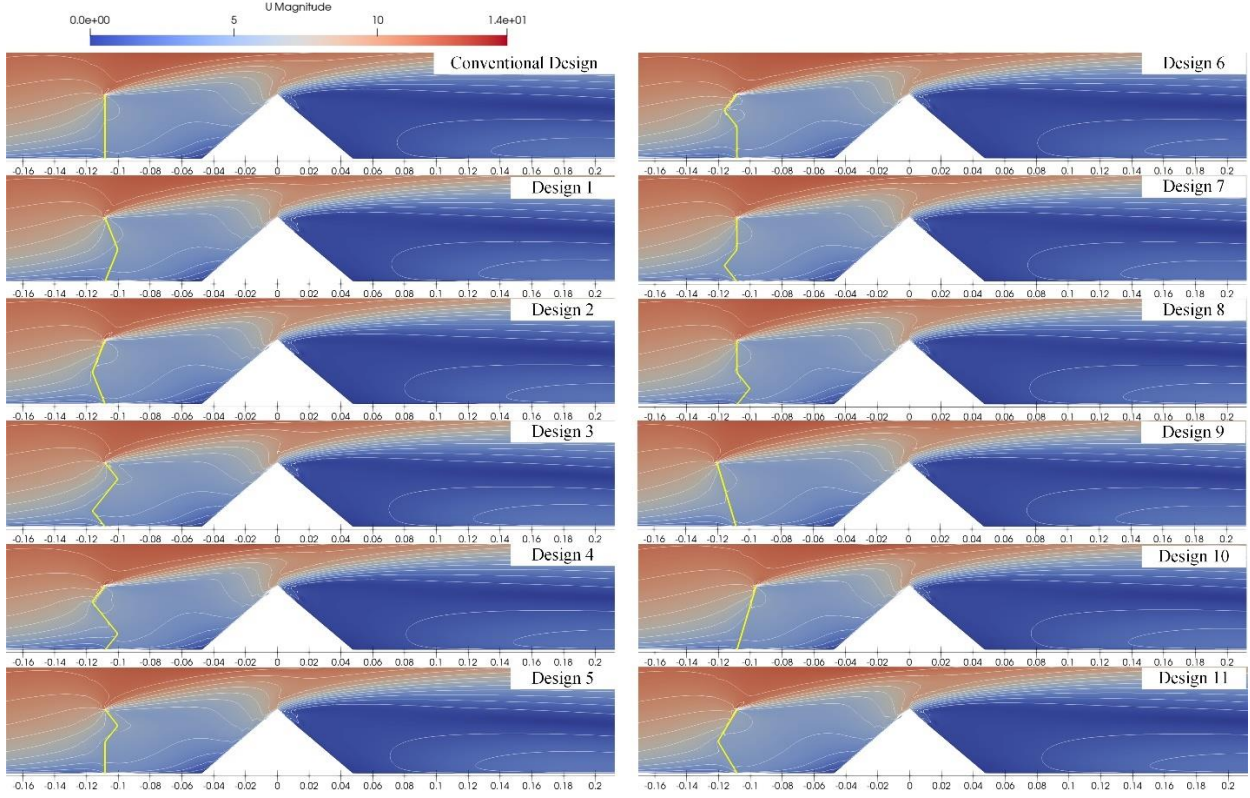

**Fig. 11.** Flow fields around the prism for different fence designs.

*3.3. Optimization of the novel design according to the sand mitigation measure (SMM)*

The optimization studies showed that design 11, which will be referred to as the new design, achieved the best reduction in both pressure and friction coefficients over the prism. Another objective of optimizing the fence design is to prevent the erosion mechanism for the conservation of fine particles and reduce wind erosion in arid regions. Therefore, the following objective function is defined for optimizing the new design in terms of erosion on the pile:

$$Minimize: F_3(\phi) = (|(\tau_x)_i|_{max}/\tau_t) \quad (20)$$

where, $|(\tau_x)_i|_{max}$ is the maximum absolute value of the local shear stress along the pile and $\tau_t$ is the threshold shear stress for the median sand diameter *d*. In this study, local shear stress in the horizontal direction is considered in the calculation of the objective function. Optimization studies are performed for different sand diameters (*d*=0.2 mm and *d*=0.5 mm) and reference velocities ($U_{ref} = 10$, 14.5 and 20 m/s) to demonstrate the practical implementation of the present approach. Corresponding threshold shear stresses are $\tau_t$ =0.09 Pa and $\tau_t$ =0.21 Pa for *d*=0.2 mm and *d*=0.5 mm, respectively [43]. Here, the new design is optimized according to the objective function (Eq. 20), and variations of the objective function with the porosity are plotted for different site conditions in Fig. 12. Simulation results show that the optimum porosity of the new design falls in a relatively narrow zone between 0.1 and 0.12.

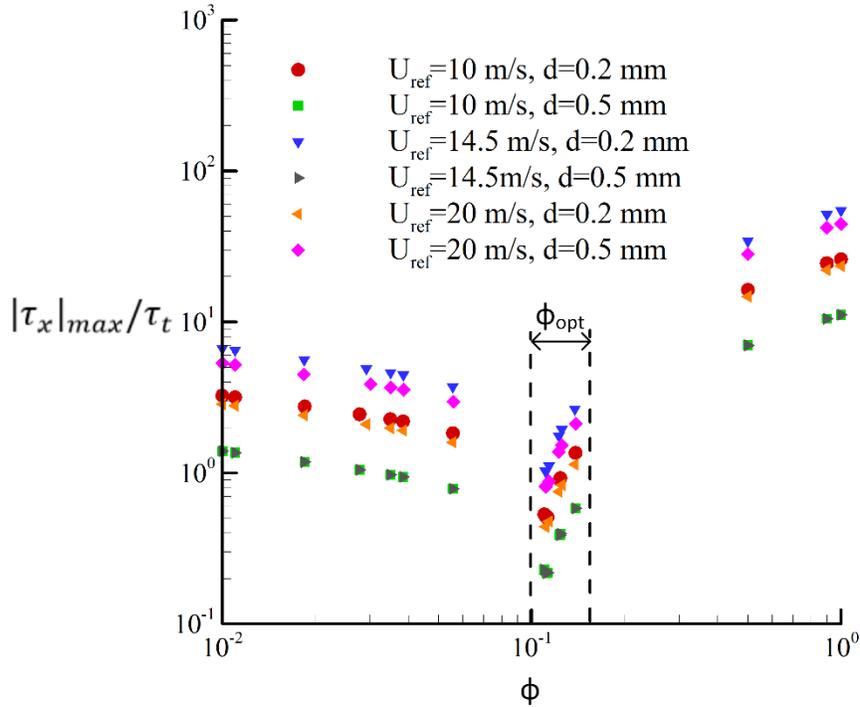

**Fig. 12.** Optimum porosity of the new design for different site conditions.

Fig. 13 compares the variations of the ratio between the maximum shear stress and threshold shear stress along the pile for the optimized new design and no fence under different conditions. The optimized new design successfully reduces SMM below 1, except for the extreme case of $U_{ref}$ = 20 m/s and $d = 0.2$ mm, which experiences severe reference velocity and small threshold shear stress. This simulation-optimization framework can be practically utilized in engineering projects to optimize the new fence design and prevent erosion processes, depending on atmospheric conditions and the average sand diameter.

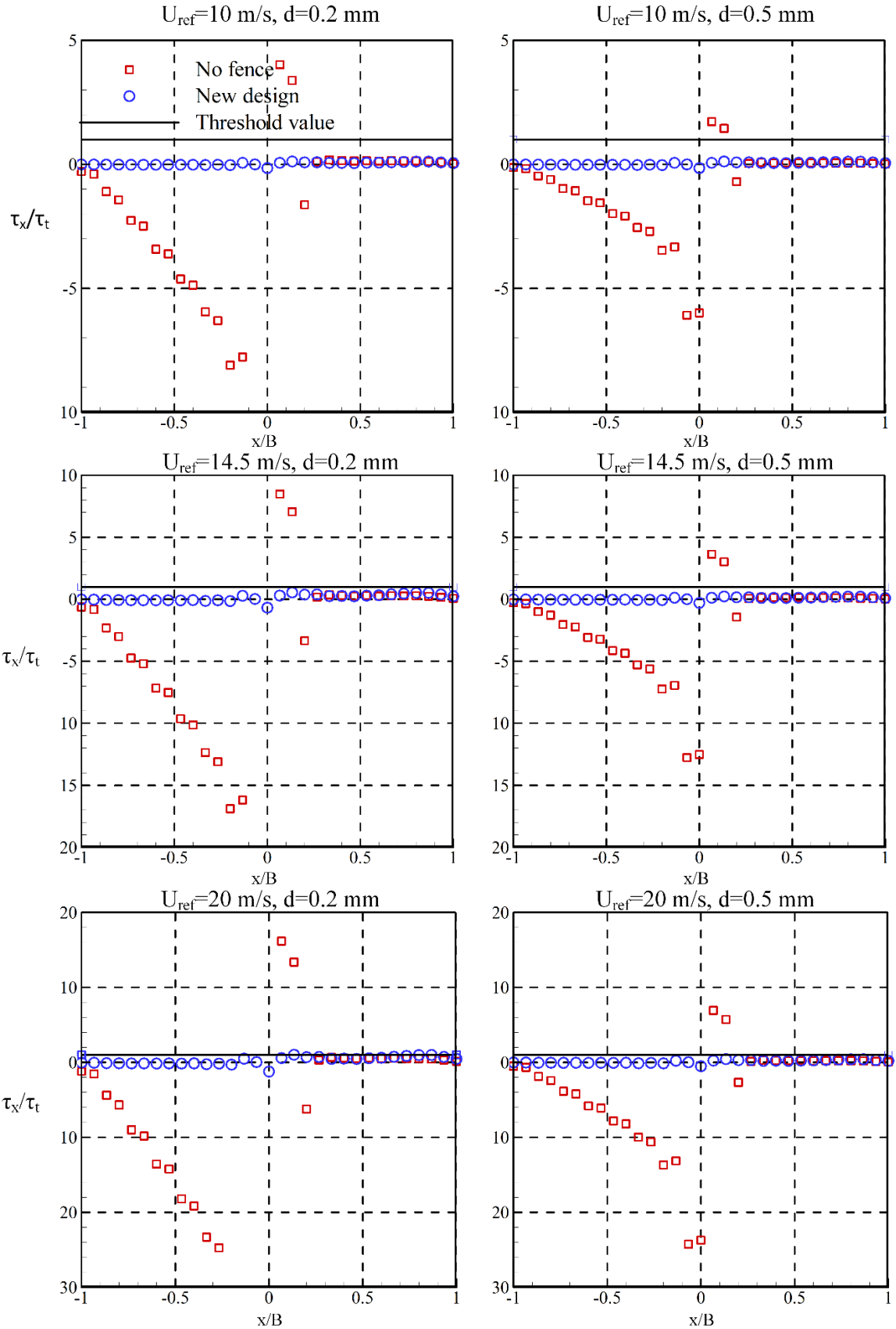

**Fig. 13.** Performance of the new design in terms of the erosion process under different site conditions.

To reduce the SMM below unity for $U_{ref} = 20$ m/s and $d = 0.2$ mm, the height of the fence is optimized using the optimum porosity in Fig. 14. The present simulation-optimization framework is slightly modified to include the optimization of the fence height for a specific porosity. The variation of the ratio with the fence height in Fig. 14 shows that the SMM can be reduced from 1.84 to 0.48 when the height of the fence is increased from 40 mm to 47 mm. Fence heights greater than the optimum height increase the SMM, which proves the existence of an optimal fence height for a specific porosity. As seen in Fig. 14, the optimized fence design still has the capacity to prevent erosion even in more extreme wind conditions.

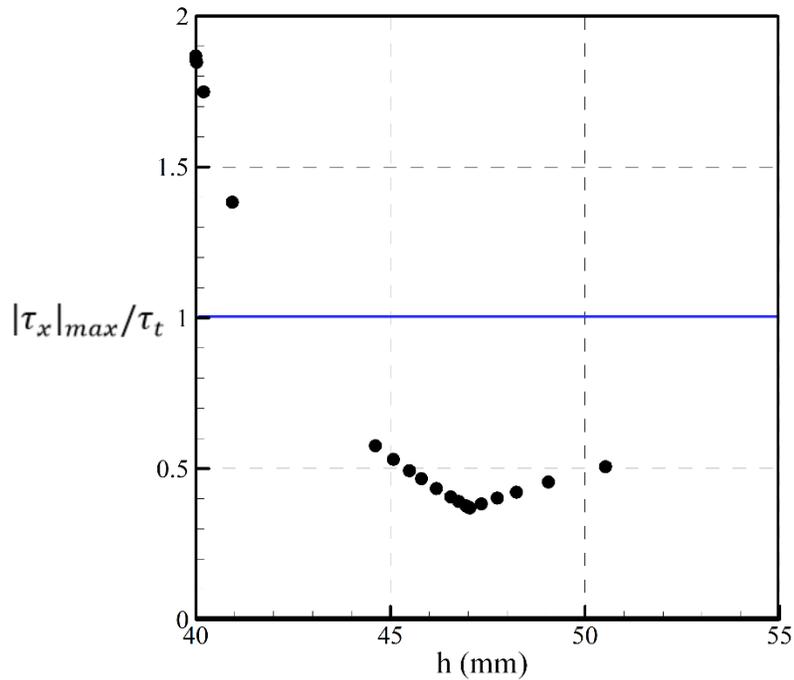

**Fig. 14.** Optimization of the height of the fence according to the SMM.

Streamlines and velocity vectors around the optimized fence design are illustrated in Fig. 15. The sharp shape of the fence deflects most of the momentum of the incident flow to the outer region of the pile. The pressure drop induced by the fence reduces the momentum of the flow significantly, resulting in a low shear stress distribution over the windward face of the pile. The recirculation zone observed downstream of the pile can also reduce the erosion process over the leeward face of the pile.

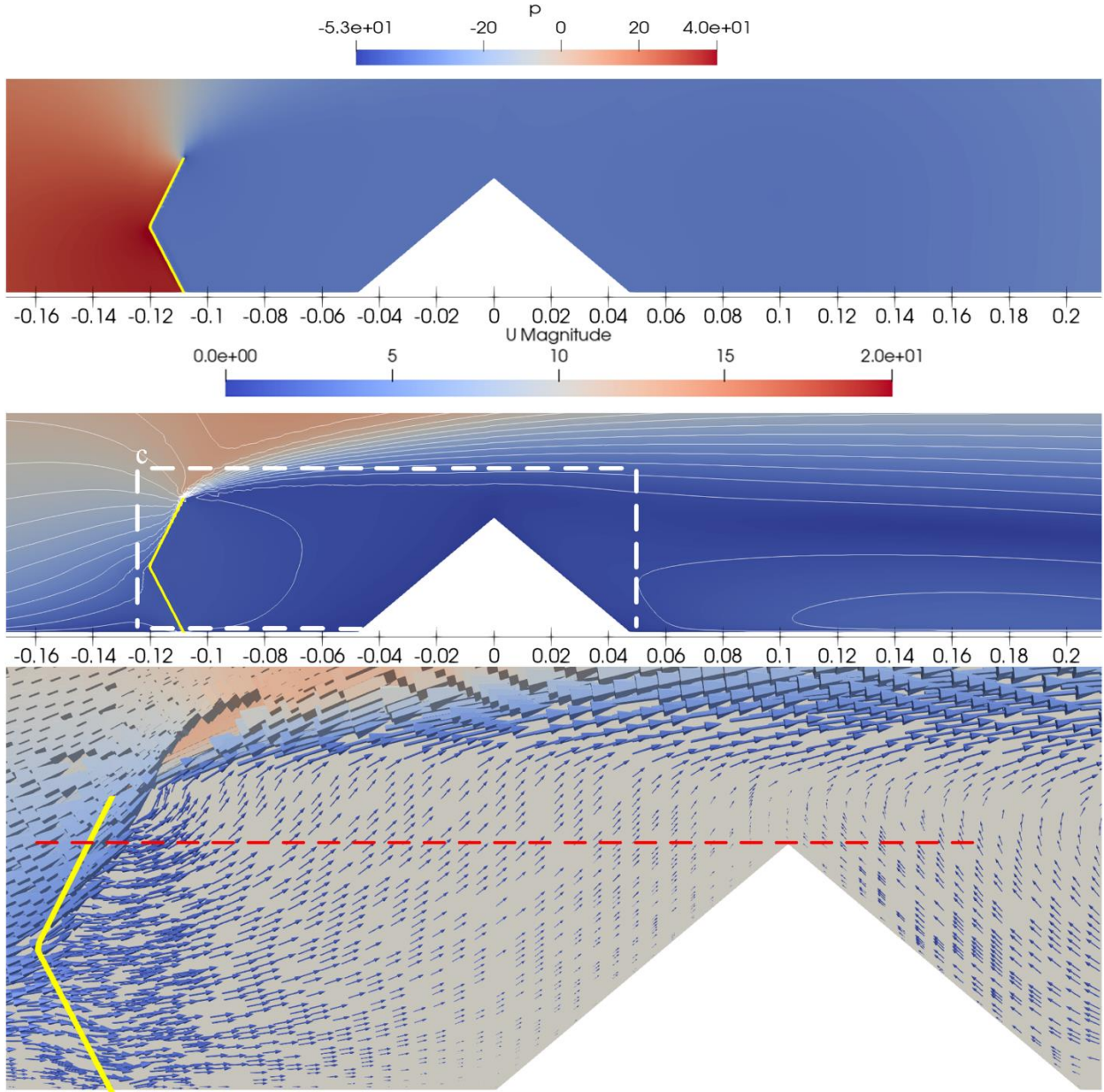

**Fig. 15.** Pressure distribution, streamlines and velocity vectors (from top to bottom) around the prism for the optimal new design ($\phi = 0.11$ and *h*=47 mm)

*3.4. Three-dimensional flow analysis of the optimal designs*

Three-dimensional unsteady numerical simulations are performed for the optimized cases to capture flow characteristics and vortex structures forming near the prism. Dimensions of the three-dimensional domain are selected the same as in the numerical study conducted by Chen et al. [22]. In this study, 20 cells are used in the z direction keeping the same mesh structure as in the two-dimensional studies and the resultant mesh consists of about one million cells in the computational domain shown in Fig. 16. Symmetry boundary condition is applied at the top and lateral boundaries to reduce the domain size and computing time for the unsteady simulations. Time derivatives of the variables are included to the governing equations for the unsteady incompressible flow. A second order-accurate Euler method is used for the discretization of the unsteady terms in the governing equations. The time step is adjusted according to the Courant number for the stability of the unsteady solution of the governing equations using the PIMPLE (pressure implicit with splitting of operator) algorithm in OpenFOAM. The Courant number is set to 5 and the calculated time step size is observed to vary around 0.0002 s during simulations. Note that the pimpleFOAM solver can yield stable results for the Courant number up to 6. Numerical simulations are performed for 70 s and time-averaging is commenced at t=50 to avoid unphysical effects originating from the initial conditions on the time-averaging.

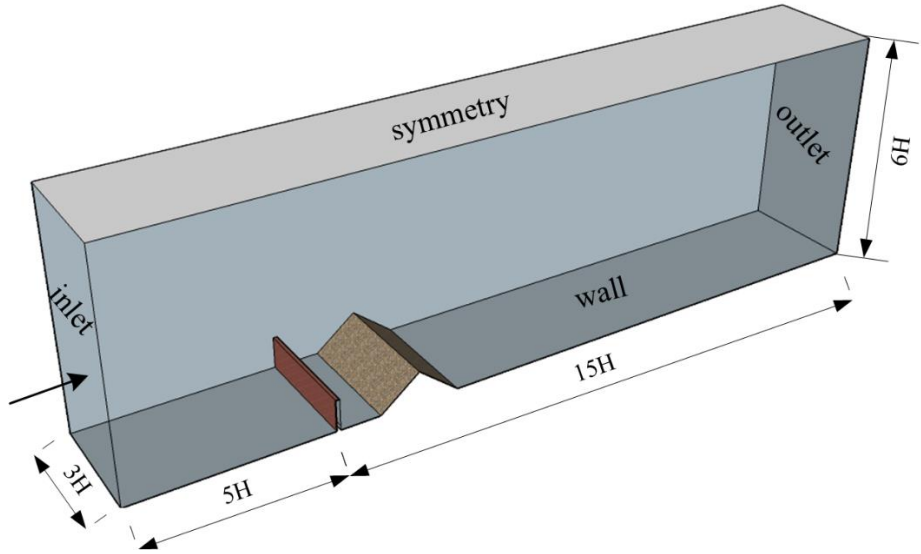

**Fig. 16.** Three-dimensional computational domain and dimensions.

Three URANS simulations are performed under the reference velocity of 14.5 m/s for the conventional fence design with the optimal porosity, proposed fence design with the optimal porosity and the proposed fence design with the optimal porosity and height. Three-dimensional streamlines and vorticity field ($\omega$) are calculated based on the mean velocity field $U_{mean}$ and visualized in Fig. 17 coloring by the magnitude of the local mean velocity $|U_{mean}|$ and vorticity $|\omega_{mean}|$ for the streamlines and vorticity field, respectively. Streamlines remain parallel to the vertical plane since geometry of each design varies only x and y directions. However, small recirculation zone observed near the toe of the proposed fence needs to be taken into account for the prevention of the local erosion. As can be seen from the streamlines in Fig. 17b and c, the convergent flow was successfully deflected to the crest of the prism with the impingement of the flow emerging from the lower inclined part of the proposed design. On the other hand, the recirculation zone downstream of the fence tends to move in the flow direction when the height of the proposed design is optimized (Fig. 17c), which brings an additional erosion mitigation performance. Vorticity fields shown on the right hand side of Fig. 17 clearly show that the strong vortex emanating from the conventional fence approaches the prism creating tangential effects on the prism, which increases the erosion risk on the prism. The large vortex is broken into small eddies and damped at the windward of the prism and the damped vortex is directed to the outer region of the prism with the application of the proposed fence design. Comparison of Fig. 17 b and c confirms that the optimization of the fence height resulted in a less elongated vortex near the crest of the leeward side of the prism, which reduces the SMM under severe wind conditions.

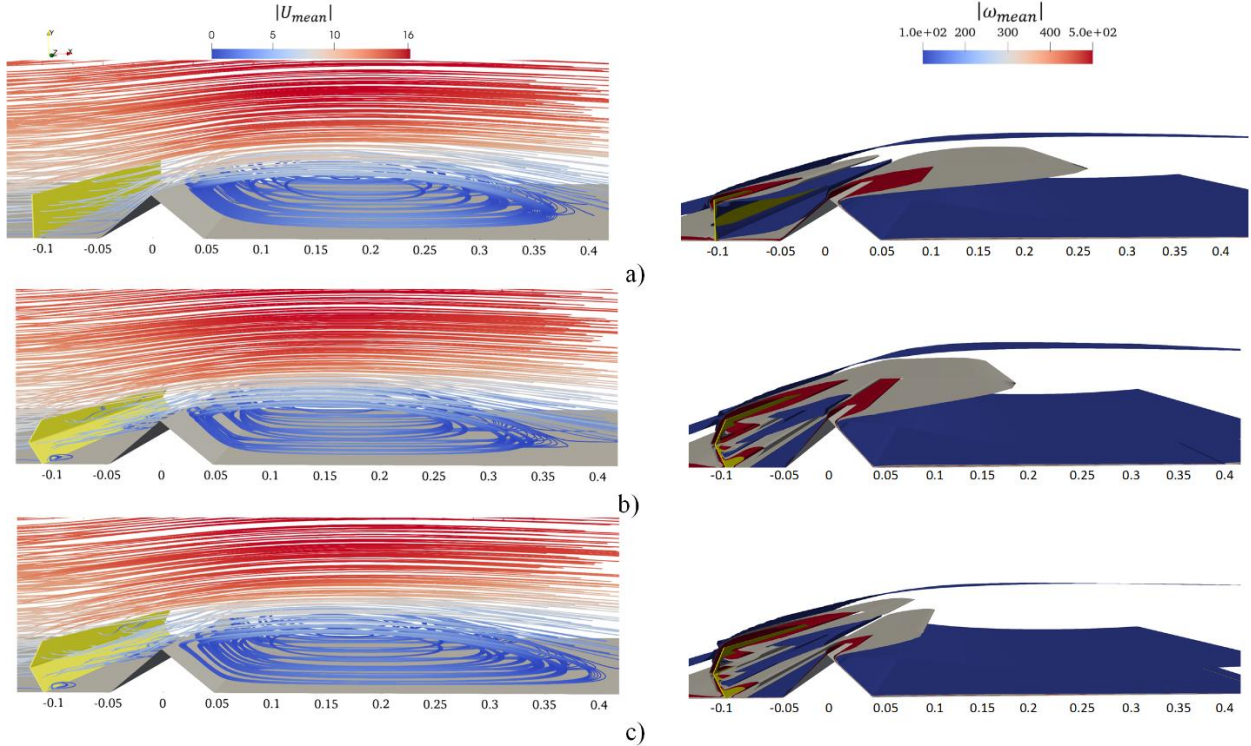

**Fig. 17.** Visualization of the three-dimensional streamlines and vorticity fields for the a) conventional fence design with optimal porosity, b) proposed fence design with optimal porosity and c) proposed fence design with optimal porosity and height.

## 4. Conclusions

In this work, we have developed and employed an open-source simulation-optimization framework for optimizing a porous fence under atmospheric conditions. The simulation results have been validated with previously reported experimental measurements on the prism, both with and without a fence of different porosities. The porosity of the conventional design has been optimized using the proposed simulation-optimization toolbox to reduce wind erosion over a triangular prism. The FRCG method has been found to converge to an optimal solution with fewer iterations than the full Newton and DIRECT methods to minimize pressure forces over the triangular prism. Performance improvements for various fence shapes have been quantified in comparison to the no-fence case. Among the 11 designs tested in the present study, design 2 and design 11 have been found to be superior to the conventional design in terms of pressure and friction reductions. The porosity of design 11 has been optimized according to the SMM to avoid the erosion process under different reference velocities and sand grain diameters. The proposed fence design has been able to successfully reduce the SMM below unity except for the reference velocity of 20 m/s and sand diameter of 0.2 mm, which is an extreme case in wind engineering applications. The fence height has been optimized using the present simulation-optimization framework to prevent the onset of wind-induced erosion for this case. Unsteady three-dimensional simulations are performed for the fences optimized according to the porosity and height. Three dimensional streamlines and vorticity fields revealed that the large vortex emerging from the fence were broken into small eddies and deflected to the outer region of the fence with the effect of the flow emerging from the lower inclined part of the proposed fence design. Optimization of the fence height resulted in a less elongated vortex near the crest of the leeward side of the prism, which contributes reducing the SMM under severe wind conditions. Consequently, the porosity and height of the proposed fence design can be optimized using the present simulation-optimization framework depending on the wind conditions and sand diameter in engineering projects.

**Supplementary Material**

See the supplementary material for OpenFOAM and Dakota files for the simulation and optimization of the fence designs. The material is publicly available in Zenodo at https://doi.org/10.5281/zenodo.7698203, Ref. 44.

**Acknowledgment**

The numerical calculations reported in this paper were fully performed at TUBITAK ULAKBIM, High Performance and Grid Computing Center (TRUBA resources). The authors express their gratitude to the referees for their valuable and constructive criticism, which significantly enhanced the quality of the manuscript.

**Funding**

Not applicable.

**Contributions**

Adnan Tolga Kurumus: Formal analysis, Resources, Validation, Investigation, Data Curation, Writing—Original Draft, Visualization, Nazhmiddin Nasyrlayev: Investigation, Data Curation, Formal analysis, Resources, Writing—Original Draft, Writing—Review & Editing, Visualization, Ender Demirel: Supervision, Writing—review & editing, Conceptualization, Methodology.

**Conflict of interest**

The authors declare no competing interests for this study.